\documentclass[floatfix,lengthcheck,showpacs,amssymb,amsmath,amsfonts,twocolumn]{revtex4-2}

\usepackage{lineno}
\usepackage[utf8]{inputenc}
\usepackage{color}
\usepackage{graphicx}
\usepackage{float}
\usepackage{subfigure}
\usepackage{amsmath}
\usepackage{multirow}
\usepackage{tikz}
\usetikzlibrary{arrows,shapes,trees,decorations.pathreplacing}
\usepackage{import}
\usepackage{svn-multi}
\usepackage{hyperref}
\usepackage{gensymb}
\hypersetup{
    colorlinks=true,
    linkcolor=blue,
    filecolor=magenta,      
    urlcolor=cyan,
}

\newcommand{\msun}{\ensuremath{\mathrm{M}_{\odot}}}

\usepackage{acro}
\DeclareAcronym{GR}{
	short = GR,
	long  = general relativity
	}
\DeclareAcronym{SNR}{
	short = SNR,
	long  = signal-to-noise ratio
	}
	
\DeclareAcronym{BBH}{
	short = BBH,
	long  = binary black hole
	}
	
\begin{document}

\title{Binary black hole spectroscopy: a no-hair test of GW190814 and GW190412}

\author{Collin D. Capano}
\email{collin.capano@aei.mpg.de}
\author{Alexander H. Nitz}
\affiliation{Max-Planck-Institut f{\"u}r Gravitationsphysik (Albert-Einstein-Institut), D-30167 Hannover, Germany}
\affiliation{Leibniz Universit{\"a}t Hannover, D-30167 Hannover, Germany}

\begin{abstract} 
Gravitational waves provide a window to probe general relativity (GR) under extreme conditions. The recent observations of GW190412 and GW190814 are unique high-mass-ratio mergers that enable the observation of gravitational-wave harmonics beyond the dominant $(\ell, m) = (2, 2)$ mode. Using these events, we search for physics beyond GR by allowing the source parameters measured from the sub-dominant harmonics to deviate from that of the dominant mode. All results are consistent with GR. We constrain the chirp mass as measured by the $(\ell, m) = (3, 3)$ mode to be within $0_{-3}^{+5}\%$ of the dominant mode when we allow both the masses and spins of the sub-dominant modes to deviate. If we allow only the mass parameters to deviate, we constrain the chirp mass of the $(3, 3)$ mode to be within $\pm1\%$ of the expected value from GR.

\end{abstract}

\date{\today}

\maketitle

\begin{figure*}
    \centering
    \includegraphics[width=2.0\columnwidth]{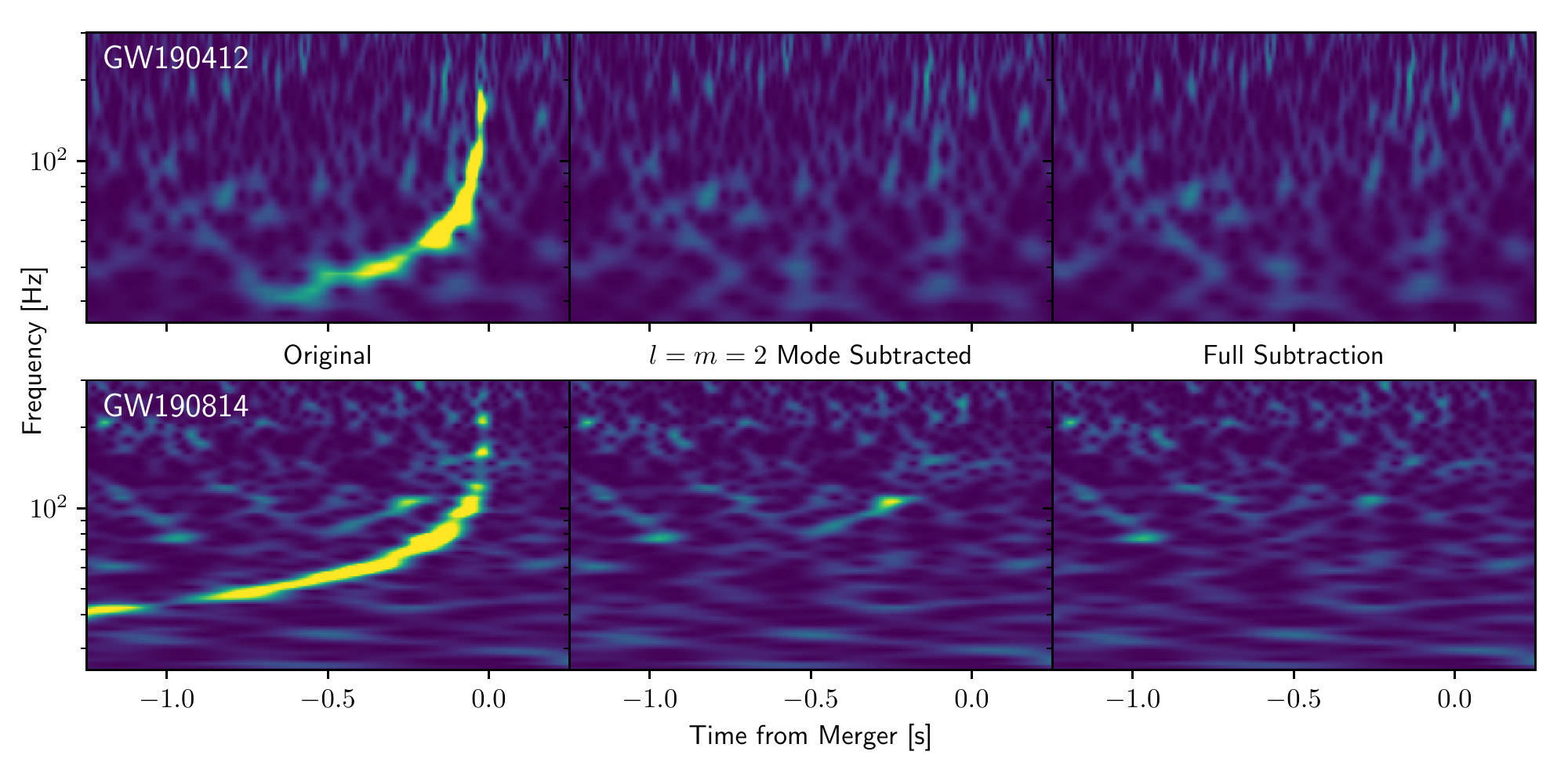}
    \caption{Time-frequency plot of the coherently-summed data around GW190412 (top) and GW190814 (bottom). The unaltered data (left) is shown along with the residuals after subtracting only the $(\ell, m) = (2,2)$ mode (center) or the full signal model (right) at the maximum-likelihood parameters. Faint visual evidence of the $(\ell, m) = (3,3)$ mode is apparent for GW190814, which has SNR $\sim7$ and roughly parallels the clear $(2, 2)$ mode, but at $1.5\times$ higher frequency.}
\label{fig:data}
\end{figure*}

\section{Introduction}

Advanced LIGO~\cite{TheLIGOScientific:2014jea} and Virgo~\cite{TheVirgo:2014hva} have detected more than a dozen binary black hole mergers to date~\cite{Nitz:2018imz,Nitz:2019hdf,Nitz:2020naa,Venumadhav:2019tad,Venumadhav:2019lyq,Zackay:2019btq,LIGOScientific:2018mvr}, with dozens of additional candidates reported during the recently concluded third observing run~\cite{GraceDBo3page}. The most recent two binary black hole detections, GW190412~\cite{LIGOScientific:2020stg} and GW190814~\cite{Abbott:2020khf} were found to have unusually asymmetric masses, with mass ratio $\sim$3 and 9, respectively. High-mass-ratio mergers such as these provide new insights into binary black hole formation channels~\cite{Olejak:2020oel,Rodriguez:2020viw,Hamers:2020huo,Gerosa:2020bjb,Safarzadeh:2020qrc,Kimball:2020opk}. GW190814, in particular, has sparked considerable interest. Its lighter component object --- which had a mass of $\sim2.6\,\msun$ --- is within the hypothesized lower ``mass gap''~\cite{Abbott:2020khf,Farr:2010tu,Bailyn_1998,Ozel:2010su,Ozel:2012ax}, challenging existing formation models~\cite{Zevin:2020gma, Broadhurst:2020cvm, Safarzadeh:2020ntc, Zhang:2020zsc, Mandel:2020cig, Yang:2020xyi,Godzieba:2020tjn,Dexheimer:2020rlp, Kinugawa:2020tbg, Rastello:2020sru}.

In addition to providing insights into stellar evolution, the direct detection of binary black holes with gravitational waves has provided new opportunities to test general relativity (GR) in the strong-field regime~\cite{TheLIGOScientific:2016src,Yunes:2016jcc,Krishnendu:2019tjp,Wang:2020cub,Wang:2020pgu,Abbott:2017vtc,LIGOScientific:2019fpa,Nielsen:2019ekf,Carullo:2018gah,Haster:2020nxf,Liu:2020slm,Shao:2020shv,Cabero:2017avf,Roy:2019phx,Abbott:2018lct}. One of the most exciting (and elusive) possibilities in this new era is a test of the no-hair theorem~\cite{Ota:2019bzl,Forteza:2020hbw,Maselli:2017kvl,Isi:2019aib,Bhagwat:2019dtm,Cabero:2019zyt,Dreyer:2003bv,Maselli:2019mjd,Bhagwat:2019bwv}. The no-hair theorem states that all stationary black holes are entirely characterized by three externally observable parameters: the object's mass, spin, and charge~\cite{PhysRevLett.26.331,PhysRev.164.1776}. This is reduced to just mass and spin for astrophysical black holes, as it is difficult for them to accumulate any appreciable charge. A compact object that requires more than two parameters to characterize it is therefore not a black hole as described by GR.

Although the no-hair theorem is a statement about stationary black holes, a perturbed black hole will radiate gravitational waves, thereby asymptoting to a Kerr spacetime. For small perturbations, the emitted gravitational wave is a superposition of quasi-normal modes (QNM)~\cite{Vishveshwara:1970zz, Press:1971wr, Teukolsky:1973ha, Chandrasekhar:1975zza}. As a consequence of the no-hair theorem, the frequency and damping times of these modes are uniquely defined by the black hole's mass and spin. This suggests a test~\cite{Dreyer:2003bv}: infer the mass and spin from each mode separately, then compare the estimates. A discrepancy would imply a violation of the no-hair theorem. Several studies have investigated applying this test (known as \emph{black hole spectroscopy}) to the final black hole that is formed after a binary black hole merger~\cite{Dreyer:2003bv,Berti:2005ys,Berti:2007zu,Gossan:2011ha,Berti:2016lat,Giesler:2019uxc,Cabero:2019zyt,Islam:2019dmk}. 

Performing black hole spectroscopy on a binary merger remnant is challenging. The post-merger waveform damps away quickly [in $\mathcal{O}$(ms) for stellar-mass black holes], and the amplitudes of the sub-dominant modes are (at best) $\lesssim 30\%$ of the dominant mode \cite{Borhanian:2019kxt}. The signal-to-noise ratio (SNR) of the pertinent signal is therefore relatively small. Furthermore, \emph{when} the post-merger perturbation is sufficiently small such that linear-perturbation theory can be applied is a matter of debate. Many studies have found that it is necessary to wait at least $10M$ after the merger if only fundamental modes are used in the waveform model~\cite{Buonanno:2006ui,Berti:2007fi,TheLIGOScientific:2016src,Bhagwat:2017tkm}. However, it has been shown that the signal immediately after merger can be modelled as a superposition of QNMs if overtones of the dominant mode are included~\cite{Leaver:1986gd,Buonanno:2006ui,Baibhav:2017jhs,Giesler:2019uxc}. This substantially increases the SNR available for spectroscopy, but introduces additional technical and conceptual issues that are still being resolved~\cite{Berti:2006wq,Zhang:2013ksa,Forteza:2020hbw,Bhagwat:2019dtm,Okounkova:2020vwu,Pook-Kolb:2020jlr}. Even so, the best constraint to date from that approach (obtained on the frequency of the loudest overtone) is only $\sim \pm 40\%$ of the expected GR value ($90\%$ credible interval)~\cite{Isi:2019aib}.

Here we take a different tack: we perform \emph{binary} black hole spectroscopy. We use the entire observable signal to search for hints of non-GR degrees of freedom, which we generically refer to as ``hair''. The gravitational wave emitted throughout the inspiral and merger of a binary black hole system can be decomposed into a superposition of spin-weighted spherical harmonics. Assuming circular orbits and vacuum spacetime, all of these harmonics should be dependent on just eight ``intrinsic'' parameters --- the two components' masses and the magnitude and relative orientation of their spins. This is because the initial black holes are very nearly Kerr when far apart. Therefore, as with traditional black hole spectroscopy, we can construct a test of the no-hair theorem by independently measuring the intrinsic parameters from each gravitational-wave harmonic. If the parameters do not agree across harmonics, then this suggests the presence of hair in the system, a potential violation of GR.

We apply this test to GW190412 and GW190814. Due to its low mass, uncertainty exists as to whether the lighter object in GW190814 is a neutron star or a black hole~\cite{Abbott:2020khf}. Here, we assume it is a black hole. This is reasonable given that observations of the binary neutron star merger GW170817 disfavor neutron-star equations of state that support masses $\gtrsim 2.3\,\mathrm{M}_\odot$ \cite{Shibata:2019ctb,Margalit:2017dij}. With SNRs of 3.5 and 7, respectively, in their $(\ell, m) = (3, 3)$ mode, GW190412 and GW190814 are the only two detections to date that have a measureable sub-dominant mode. Figure~\ref{fig:data} shows the data surrounding both of these mergers. 

Islam et al.~\cite{Islam:2019dmk} and Dhanpal et al.~\cite{Dhanpal:2018ufk} proposed a similar test as what we preform here. The authors of those publications investigated allowing the chirp mass and mass ratio to vary from the $\ell = 2$ mode, using common deviation parameters for all sub-dominant harmonics. They applied that test to a set of simulated signals from numerical relativity, finding that the chirp mass deviation of the sub-dominant modes could be constrained to less than a percent at signal-to-noise ratio 25. Our test here differs in that we allow both spins and masses to vary independently for every harmonic that we consider, along with the phase. We make these choices because our aim is to perform an agnostic test on the initial black holes, similar to what is done on the final remnant in traditional black hole spectroscopy. However, we also perform a more constrained test in which only the masses and phase of the sub-dominant modes are varied, keeping the spins fixed to their GR values. This is more similar to the test proposed in Refs.~\cite{Islam:2019dmk} and \cite{Dhanpal:2018ufk}. As discussed below, we obtain constraints that are consistent with what those publications found using simulated GR signals.

\begin{figure*}[ht!]
    \includegraphics[width=2.0\columnwidth]{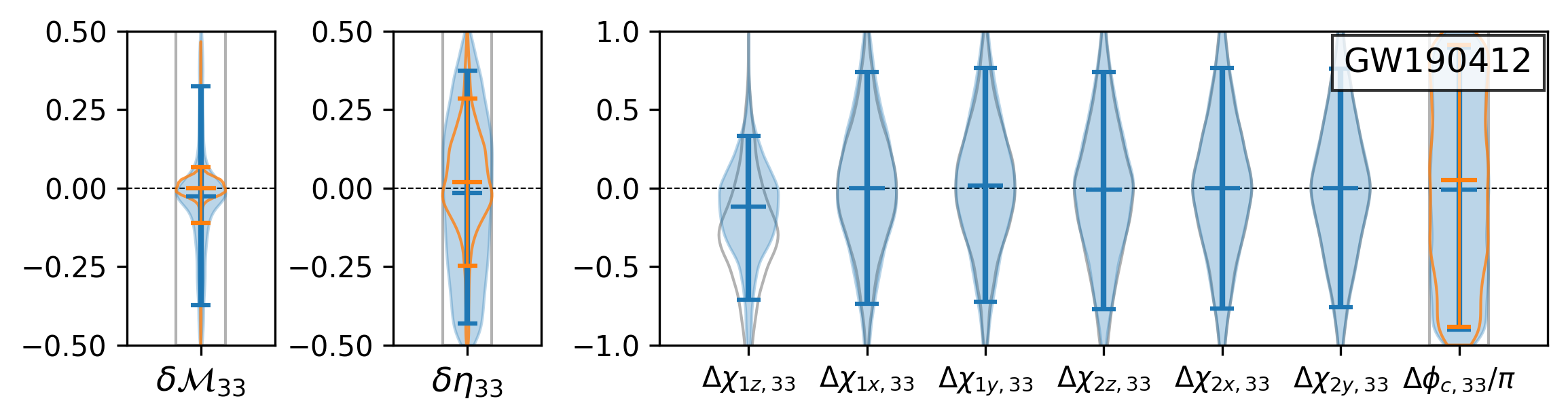} \\
    \includegraphics[width=2.0\columnwidth]{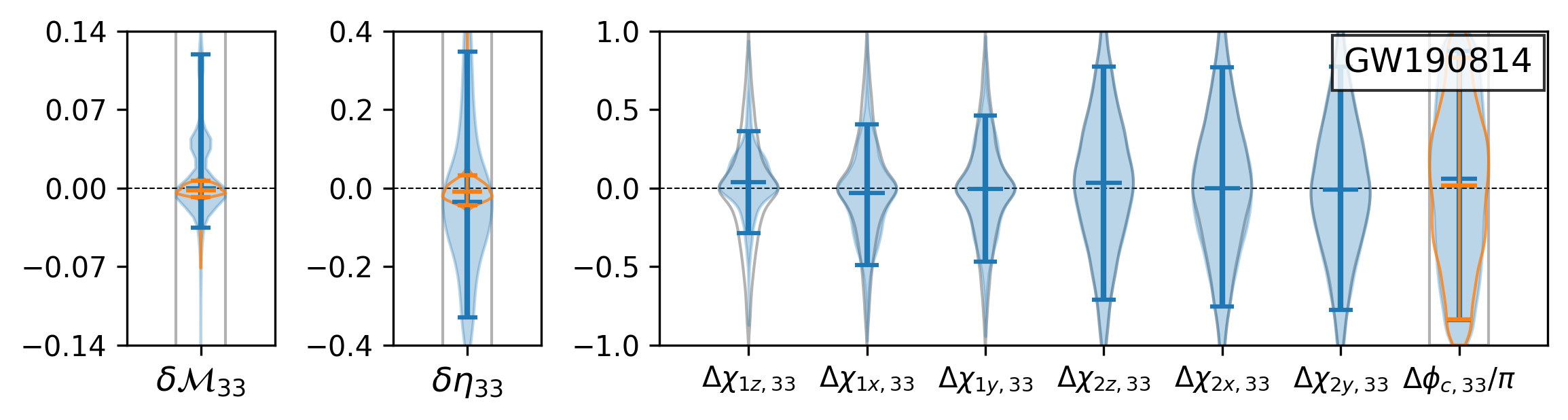}
    \caption{Marginal posterior distributions of the $(\ell, m) = (3, 3)$ deviation parameters for GW190412 (top) and GW190814 (bottom). Blue regions/lines are from the analysis in which we allow both the masses and spins to deviate. Orange lines show the same result when the sub-dominant mode spins are fixed to the dominant-mode value. Horizontal hashes indicate the median (center) and 90\% credible regions. Gray lines show the marginalized prior distributions. Since the prior distributions on the $\Delta \chi_{ik, \ell m}$ are dependent on the dominant-mode spins $\chi_{ik, 22}$, we show spin priors conditioned on the $\chi_{ik, 22}$ posteriors.}
\label{fig:violin}
\end{figure*}

\section{Binary black hole spectroscopy}

The full gravitational wave as seen by an external observer at distance $D_L$ from the source can be decomposed into a spin-weighted spherical harmonic basis,
\begin{align}
\label{eqn:gw}
h_{(+,\times)} &= \nonumber \\
\frac{1}{D_L} \left(\Re, -\Im\right) &\sum_{\ell m} {}_{-2}Y_{\ell m}(\iota, \phi) A_{\ell m}(\mathbf{\Theta}) e^{i(\Psi_{\ell m}(\mathbf{\Theta}) + m\phi_c)}.
\end{align}
Here, $h_{(+\,\times)}$ is the ``plus'' and ``cross'' polarization of the gravitational wave, the inclination $\iota$ is the angle between the line-of-sight of the observer and the $z$-axis of the center-of-mass frame of the source, and $\phi$ is the azimuthal angle of the observer with respect to this frame. The source frame is oriented such that the $z$-axis is aligned with the orbital angular momentum at some fiducial reference time (here, chosen as the time that the frequency of the dominant-mode is $20\,$Hz); $\phi_c$ is the phase of the gravitational wave at that time.

The set $\mathbf{\Theta}$ represents all of the intrinsic parameters describing the source binary. Unless the binary was formed by a recent dynamical capture, the orbit circularizes before it enters the sensitive frequency band of the LIGO and Virgo instruments~\cite{Peters:1964zz}. Therefore, assuming circular orbits, a binary black hole is uniquely defined by eight parameters: the two component masses $m_{(1,2)}$ and the magnitude and orientation of each object's spin $\vec{\chi}_{(1,2)}$ at a reference epoch.

To perform spectroscopy on the full waveform, we allow the intrinsic parameters and the phase of the sub-dominant modes to deviate from the dominant, $(\ell, m) = (2, 2)$ mode. In other words, we replace $\mathbf{\Theta}$ and $\phi_c$ in Eq.~\eqref{eqn:gw} with $\mathbf{\Theta}_{\ell m}$ and $\phi_{c,\ell m}$, respectively. We examine two cases. In the first, we allow all intrinsic parameters (masses and spins) of the sub-dominant modes to independently diverge from the dominant mode. In the second scenario we only allow the mass parameters to differ. In both cases we allow $\phi_{c}$ to vary.

We expect any deviations from GR, if present, to be small. Since a binary's chirp mass $\mathcal{M} = (m_1 m_2)^{3/5}/(m_1+m_2)^{1/5}$ is more accurately measured than the individual component masses, we parameterize the sub-dominant masses in terms of fractional deviations from the dominant-mode chirp mass $\mathcal{M}_{22}$ and symmetric mass ratio $\eta_{22} = m_{1,22}m_{2,22}/(m_{1,22}+m_{2,22})^2$. Specifically, we define
\begin{align*}
\mathcal{M}_{\ell m} &\equiv \mathcal{M}_{22}(1 + \delta \mathcal{M}_{\ell m}),\\
\eta_{\ell m} &\equiv \eta_{22}(1 + \delta \eta_{\ell m}),
\end{align*}
and allow $\delta \mathcal{M}_{\ell m}$ and $\delta \eta_{\ell m}$ to vary uniformly between $\pm0.5$. The spins are varied independently for each mode, using the same prior (uniform in magnitude and isotropic in direction) as the dominant mode spins. We then report the absolute difference from the dominant mode for each component,
\begin{equation*}
\Delta \chi_{ik, \ell m} \equiv \chi_{ik, \ell m} - \chi_{ik, 22},
\end{equation*}
where $i = 1, 2$ and $k = x,y,z$. The absolute difference in the modes' phase $\Delta \phi_{c,\ell m} = \phi_{c,\ell m} - \phi_{c,22}$ is varied uniformly between $\pm\pi$. We use standard astrophysical priors for the remaining parameters: uniform distributions are used for comoving volume, coalescence time, phase, and the source-frame component masses $m^{\rm src}_{1,2}$; isotropic distributions are used for sky location, inclination, and polarization.

To model the gravitational-wave signal, we use the recently developed IMRPhenomXPHM waveform~\cite{Pratten:2020ceb}. This model includes the effects of both orbital precession and sub-dominant modes on the gravitational waveform, and has been tested against numerical relativity simulations. The $(\ell, m) = (3, 3)$ mode was the only measureable sub-dominant mode in both GW190412 and GW190814. Based on the estimated parameters, the next most significant sub-dominant mode in these events should be the $(\ell, m) = (2, 1)$ mode, although we expect its signal-to-noise ratio to be too small to provide any meaningful constraints. All other modes are too weak to measure. We therefore model the signals using the sum of the $(2, 2)$, $(3, 3)$, and (as a sanity check) $(2, 1)$ modes. We neglect all other modes in our analysis. To sample the parameter space, we use the open-source PyCBC Inference toolkit~\cite{Biwer:2018osg,pycbc-github} with the parallel tempered emcee-based sampler~\cite{Vousden:2015, emcee}.

\section{Results}

Constraints on the deviations of the $(\ell, m) = (3, 3)$ mode for each event are shown in Fig.~\ref{fig:violin}. We report marginalized posterior distributions on the fractional difference in chirp mass $\delta \mathcal{M}_{33}$ and symmetric mass ratio $\delta \eta_{33}$ from the $(2,2)$ mode, along with the absolute differences in the six spin components and the reference phase $\phi_c$.  All parameters are consistent with the GR values: a zero deviation is within the 90\% credible interval of every parameter.

\begin{figure}
    \centering
    \includegraphics[width=1.0\columnwidth]{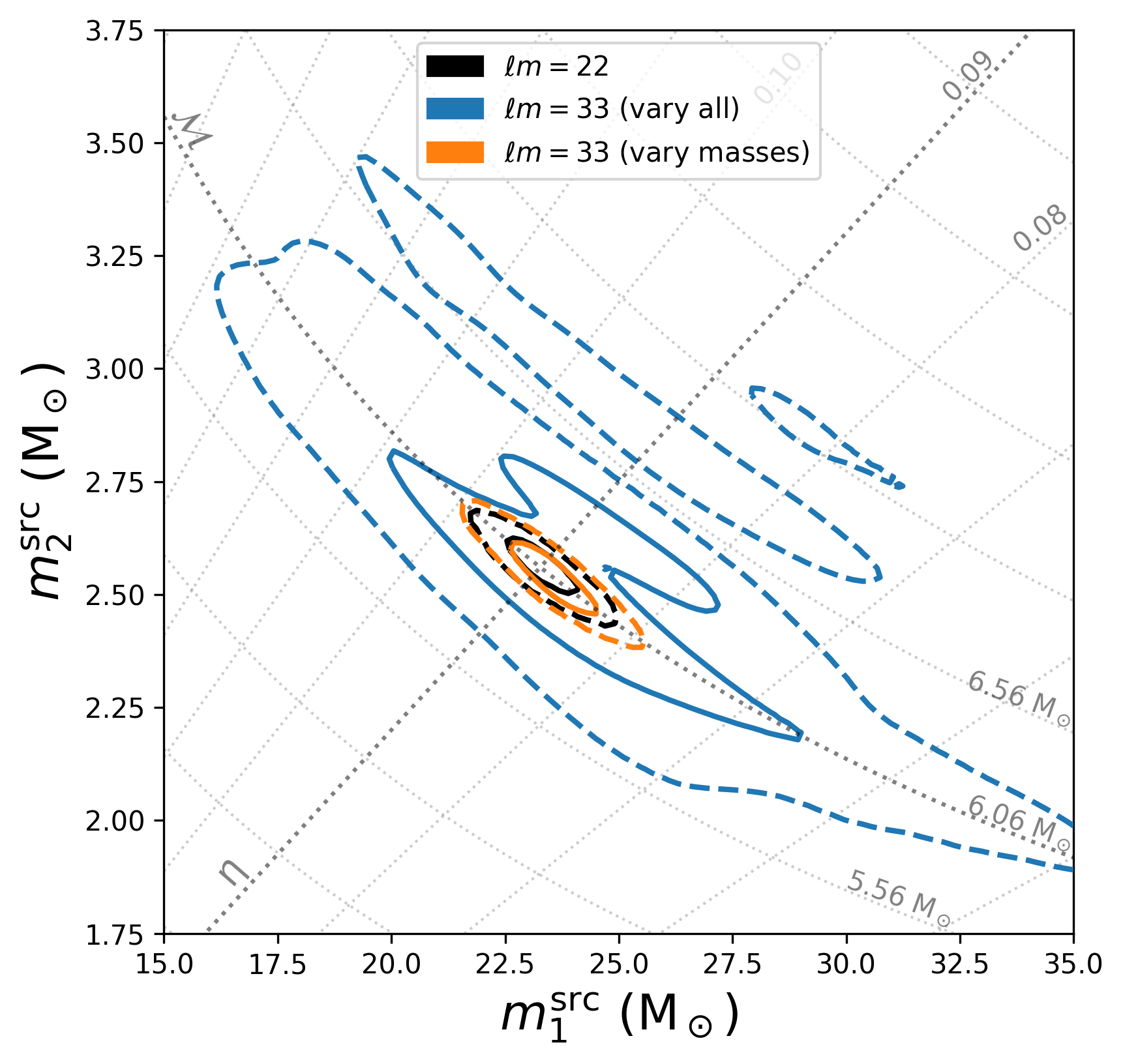}
\caption{Marginal posterior distributions on the source-frame component masses $m^{\rm src}_{1,2}$ of GW190814, as measured by the dominant [$(\ell, m) = (2, 2)$] and $(3, 3)$ modes. The contours show the 50\% (solid) and and 90\% (dashed) credible regions. Lines of constant chirp mass $\mathcal{M}$ and symmetric mass ratio $\eta$ are indicated by the gray dotted lines, with the darker lines indicating the maximum likelihood value of the $(2, 2)$ mode. Black lines show the posterior on the masses as measured by the $(2, 2)$ mode, blue lines show the $(3, 3)$ mode when spins are allowed to deviate along with the masses and phase; orange lines show the $(3, 3)$ mode when only the masses and phase are allowed to deviate.}
\label{fig:m1m2}
\end{figure}

We obtain the strongest constraints from GW190814, with its $\delta \mathcal{M}_{33}$ being the best constrained parameter overall. We also obtain non-trivial constraints on that event's $\delta \eta_{33}$. Using the posterior on $\delta \mathcal{M}_{33}$ and $\delta \eta_{33}$, we reconstruct the masses of the source-frame masses as measured by the $(3, 3)$ mode. The resulting posterior is shown in Fig.~\ref{fig:m1m2}. Consistent with GR, the posteriors are centered on the values measured by the dominant mode.

All other parameters, including all deviations on the $(2, 1)$ mode (not shown) yield negligible constraints. This is consistent with expectations from GR. The individual component spins are not well measured for either event, and the signal-to-noise ratio of the $(2, 1)$ mode for both events is expected to be too small to provide meaningful constraints. That we do not observe any constraints on these parameters lends credibility to the constraints that we do measure from the $(3,3)$ mode, and reinforces the GR nature of these events.

Our results are also consistent with predictions from Islam et al.~\cite{Islam:2019dmk}. The constraints obtained from our analysis of GW190814, in which we only allow the masses and phase of the sub-dominant modes to deviate, are the most directly comparable results to that study. The width of our $90\%$ credible interval on $\delta \mathcal{M}_{33}$ is approximately an order of magnitude larger than what Islam et al. obtained using a non-spinning simulated GR signal. A wider uncertainty is expected due to the additional degrees of freedom in our analysis: we have allowed the parameters of each mode to vary independently, we have not assumed that GW190814 is non-spinning, and we have allowed the phase of each mode to vary in addition to the masses.

\begin{figure}
    \includegraphics[width=1.0\columnwidth]{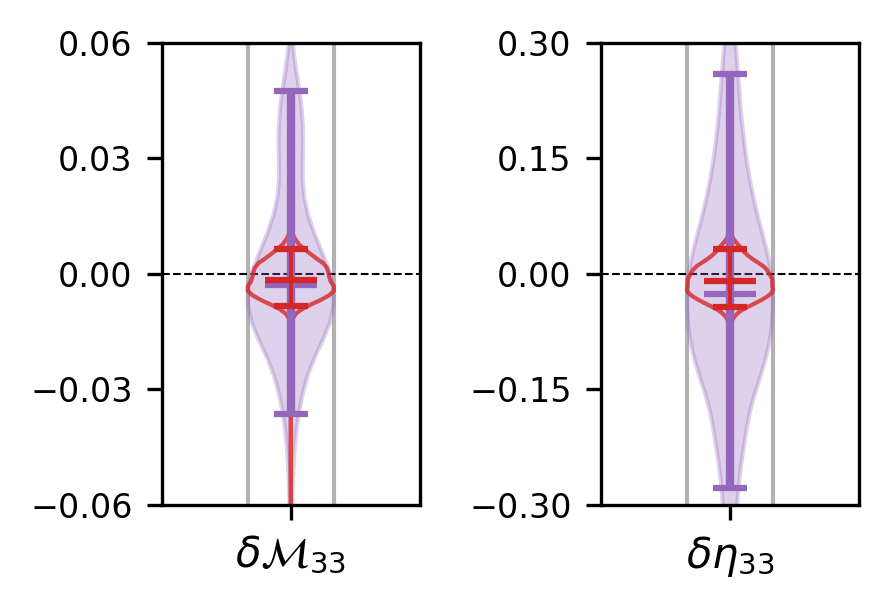}
    \caption{Combined marginal posterior distributions of $\delta \mathcal{M}_{33}$ and $\delta \eta_{33}$ from GW190412 and GW190814. Purple regions/lines are from the analysis in which we allow both the masses and spins to deviate. Red lines show the same result when we fix the sub-dominant mode spins to the dominant-mode value. Horizontal hashes indicate the median (center) and 90\% credible regions. Gray lines show the prior distribution.}
\label{fig:combined_violin}
\end{figure}

Under the assumption that the mass deviation parameters, $\delta \mathcal{M}_{33}$ and $\delta \eta_{33}$, are common between GW190412 and GW190814, we resample their likelihoods using a kernel density estimate and obtain the joint constraints shown in Fig.~\ref{fig:combined_violin}. A full summary of our analyses' marginalized constraints are shown in Table~\ref{table:numbers}.

\begin{table}
\begin{center}
    \begin{tabular}{c|c|c|c}
\hline
Event & Analysis & $\delta \mathcal{M}_{33}$ (\%) & $\delta \eta_{33}$ (\%) \\
\hline\hline
\multirow{2}{*}{GW190412}
& all & $-3^{+35}_{-35}$ & $-2^{+39}_{-41}$ \\
\cline{2-4}
& masses & $0^{+7}_{-11}$ & $2^{+27}_{-27}$ \\
\hline\hline
\multirow{2}{*}{GW190814}
& all & $0^{+12}_{-4}$ & $-4^{+38}_{-29}$ \\
\cline{2-4}
& masses & $-0.2^{+0.8}_{-0.6}$ & $-1^{+4}_{-3}$ \\
\hline\hline
\multirow{2}{*}{Combined}
& all & $0^{+5}_{-3}$ & $-3^{+29}_{-25}$ \\
\cline{2-4}
& masses & $-0.2^{+0.8}_{-0.7}$ & $-1^{+4}_{-3}$ \\ 
\hline
    \end{tabular}
\end{center}
\caption{Median and 90\% credible intervals on deviations of the $(\ell, m) = (3, 3)$ chirp mass $\delta \mathcal{M}_{33}$ and symmetric-mass ratio $\delta \eta_{33}$. These are the best-constrained deviation parameters in our analysis; other parameters are weakly constrained. We performed two analyses --- one in which the the masses, orbital phase, and spins of the sub-dominant modes are allowed to deviate from the dominant mode (``all'') and one in which only the masses and phase are allowed to deviate (``masses''). Constraints on the mass parameters are combined between the two events (``Combined'').
}
\label{table:numbers}
\end{table}

\section{Discussion}

We performed binary black hole spectroscopy on GW190814 and GW190412 to look for physics beyond general relativity. We obtained non-trivial constraints on the chirp mass and symmetric mass ratio. Consistent with GR, we find that the chirp mass as measured by the $(\ell, m) = (3, 3)$ mode and the dominant mode agree with each other to percent-level accuracy. We also combined results between GW190814 and GW190412 to further constrain deviations from GR. These results can be combined with future dectections to obtain yet tighter constraints. Over the next few years, as the LIGO and Virgo detectors reach design sensitivity, we expect the rate of mergers to increase by a factor of $5$--$10$~\cite{ObsScenario}. This should improve the limits observed here by a factor of $\sim2$--$3$, or more, if even larger mass-ratio mergers are observed.

To date, the best constraint from traditional black hole spectroscopy has been on the QNM frequency of an overtone of the $(\ell, m) = (2, 2)$ mode. That constraint, $|\delta f| \lesssim 40\%$, was obtained from an analysis of GW150914~\cite{Isi:2019aib}. Our constraint on the fractional deviation of the $(3, 3)$ chirp mass is at least an order of magnitude smaller. This is because more signal-to-noise ratio is available in the full waveform.

Is binary black hole spectroscopy a test of the no-hair theorem? Yes, or at least as much as traditional black hole spectroscopy can be said to be a test of the no-hair theorem. The primary difference is binary black hole spectroscopy probes ``long-range'' hair --- i.e., interactions that occur on an orbital length scale --- whereas black hole spectroscopy is sensitive to near-horizon effects on the scale of the remnant. Binary black hole spectroscopy is therefore a compliment to traditional black hole spectroscopy.

One complication of binary black hole spectroscopy is that the initial objects are never exactly Kerr. However, the final black hole is never truly a Kerr black hole either, only approximately so; it exists in a universe containing other matter. In that sense, \emph{neither} traditional black hole spectroscopy nor binary black hole spectroscopy can be said to be tests of the no-hair theorem, since no astrophysical black hole exists in a purely stationary spacetime. Both are null tests. A null result, as we obtained here, means that the objects in the binary are consistent with Kerr black holes, which satisfy the no-hair theorem.

Even if a non-zero deviation parameter is detected by one of these tests in a future observation, it would not necessarily mean that GR is violated. Any number of more mundane effects --- modelling uncertainties, wrong assumptions, even unexpected noise features in the detector data --- may offer an explanation. For these reasons, both traditional and binary black hole spectroscopy are perhaps better described as ``toupee tests'' rather then no-hair theorem tests. They may detect hair, but whether that indicates a violation of GR (real hair) or other systematics (fake hair) would require further study. Our hope, both with binary and traditional black hole spectroscopy, is that these tests may one day detect hair that cannot be explained by more mundane effects upon further investigation. Such a discovery could point the way to new physics.

We make available the full posterior samples from our analyses along with the configuration files necessary to reproduce our results at \url{http://github.com/gwastro/binary-black-hole-spectroscopy}. 

\acknowledgments
 We acknowledge the Max Planck Gesellschaft. We are extremely grateful to Carsten Aulbert, Henning Fehrmann, and the computing team from AEI Hannover for their significant technical support. We thank Sebastian Khan and Frank Ohme for useful discussions with regards to gravitational-waveform models. We also thank Frans Pretorius, Luis Lehner, and Alex Nielsen for providing helpful comments on alternative theories of gravity and the no-hair theorem. This research has made use of data from the Gravitational Wave Open Science Center (https://www.gw-openscience.org), a service of LIGO Laboratory, the LIGO Scientific Collaboration and the Virgo Collaboration. LIGO is funded by the U.S. National Science Foundation. Virgo is funded by the French Centre National de Recherche Scientifique (CNRS), the Italian Istituto Nazionale della Fisica Nucleare (INFN) and the Dutch Nikhef, with contributions by Polish and Hungarian institutes.
 
\bibliography{references}

\end{document}